\documentclass[11pt]{amsart}
\usepackage{amsmath}
\usepackage{amssymb, amsfonts, amsmath, amsthm, amscd}
\usepackage{mathrsfs}
\usepackage{color, graphics}
\usepackage[all]{xy}
\usepackage{hyperref}
\usepackage{pstricks,pst-plot}
\usepackage{amsxtra,epsfig,subfigure}

\def\mS{\mathcal S}
\newtheorem{lem}{Lemma}[section]
\newtheorem{thm}[lem]{Theorem}

\newtheorem{defn}[lem]{Definition}
\newtheorem{Prop}[lem]{Proposition}

\newtheorem{cor}[lem]{Corollary}

\title[Phylogenetic tree constructing algorithms fit for grid computing with SVD]
      {Phylogenetic tree constructing algorithms fit for grid computing with SVD}

\author[Young Rock Kim]{Young Rock Kim$^1$}
\address{$^1$Department of Mathematics,
   Konkuk University, Seoul, 143-701, Korea}
\email{rocky777@math.snu.ac.kr}

\author[Oh-In Kwon]{Oh-In Kwon$^1$}
\address{$^1$Department of Mathematics, Konkuk University, Seoul, 143-701, Korea}
\email{oikwon@konkuk.ac.kr}

\author[Seong-Hun Paeng]{Seong-Hun Paeng$^1$}
\address{$^1$Department of Mathematics, Konkuk University, Seoul, 143-701, Korea}
\email{shpaeng@konkuk.ac.kr}

\author[Chun-Jae Park]{Chunjae Park$^2$}
\address{$^2$IIRC, Kyunghee University, Suwon, KOREA}
\email{cjpark@khu.ac.kr}

\thanks{First author was supported in part by KRF(grant No. 2005-070-C00005
        and grant No. R14-2002-007-01001-0).
        This work was supported by grant No. R01-2006-000-10047-0 from
        the Basic Research Program of the Korean Science \& Engineering
        Foundation for third author.
        This work was supported by the SRC/ERC program of MOST/KOSEF (R11-2002-103)
        for corresponding author.}



\begin{document}

\begin{abstract}
Erikkson showed that singular value decomposition(SVD) of  flattenings determined a
partition of a phylogenetic tree to be a split (\cite{Er}).
In this paper,  based on his work, we develop new  statistically
consistent algorithms 
fit for grid computing to construct a phylogenetic tree by
computing SVD of flattenings with the small fixed number of rows.
%
%
%
\end{abstract}
\maketitle
\pagestyle{plain}

\section{Introduction}

Phylogenetic analysis of a family of related nucleic acid or protein
sequences is to determine how the family could have been derived
during evolution. Assume that evolution follows a tree model with
evolution acting independently at different sites of genome. Let the
transition matrices for this model be the general Markov model which
is more general than any other in the Felsenstein hierarchy. How to
reconstruct evolutionary trees is one of the main objects in
phylogenetics.

Since statistical models are algebraic varieties, we are interested
in defining polynomials called phylogenetic invariants for
varieties. Many authors have studied phylogenetic invariants for
different models (\cite{AR}, \cite{CF}, \cite{L}, \cite{SB},
\cite{SS}). Phylogenetic invariants have been used for phylogenetic
tree reconstruction (\cite{CF1}).

Procedures for phylogenetic analysis are linked to those for
sequence alignment. We can easily organize a group of similar
sequences with a small variation into a phylogenetic tree. On the
other hand, as sequences become more different through evolutionary
change, as they can be more difficult to be aligned. A phylogenetic
analysis of very different sequences is also hard to do since there
are many possible evolutionary paths that could have been followed
to produce the observed sequence variation. To solve these
difficulties and complexities many phylogenetic analysis programs
have been invented. The main ones in use are PHYLIP (phylogenetic
inference package, \cite{F}) available from Dr. J. Felsestein and
PAUP (\cite{S}) available from Sinauer Associates, Sunderland,
Massachusetts. Nowadays these programs provide three methods for
phylogenetic analysis - Parsimony, distance, and maximum likelihood
methods - and also give many evolutionary models for sequence
variation.

Note that splits in a phylogenetic tree play an important role in
reconstructing the phylogenetic tree (\cite{St}). Recall that
Erikkson suggested a phylogenetic tree building algorithm using SVD
of flattenings in Chapter 19 (\cite{Er}) of \cite{PS1}. In that
article, he tried to build a phylogenetic tree without concerning
the notion of distance by concentrating on the phylogenetic
invariants which are given by rank conditions of flattenings. On the
other hand, he had difficulty in dealing with the phylogenetic tree
having a large number of leaves since he had to compute SVD of
flattenings of huge size. In this paper, we construct algorithms
with SVD of flattenings of fixed number of rows, i.e., 16. We will
present tree building algorithms (Algorithm 1 and Algorithm 2) in
section 3,4 using SVD to calculate how close a matrix is to be a
certain rank. In section 5, we use the program {\bf seq-gen}
(\cite{RG}) to simulate data of various lengths for the phylogenetic
tree. After that we build a phylogenetic tree using Algorithm 1,
Algorithm 2 and Neighbor joining algorithm (NJ). It turns out that
our algorithms are efficient to construct the phylogenetic tree
involving $n\ge 15$ species for DNA sequences with respect to the
numerical stability. Also we compare our algorithms to NJ using
simulated and real Encode data. Our algorithms are suitable to
construct phylogenetic trees for
general Markov models, i.e. models coming from real data.\\

\noindent Acknowledgements

\noindent We would like to thank Lior Pachter, Bernd Sturmfels and
Seth Sullivant for discussing tree building algorithm and
encouraging us to write this paper. Also first author thanks
organizers and participants of Summer school ``Algebraic statistics,
Tropical geometry, Computational biology" in Norway for warm
hospitality and helpful discussion about these areas during he
attended in the school.

\section{Notations and Preliminaries}

In this section we explain known results by Erikkson and basic
concepts in the book: Algebraic statistics for computational biology
(\cite{PS1}) for the later use. We will present basic theorem which
plays an important role in this paper.

A phylogenetic $X$-tree $T$ is a tree with leaf set $X$ and no
vertices of degree two. If every interior vertex of a $X$-tree has
degree three, then $T$ is called a trivalent tree. A split $A|B$ of
$X$ in a tree $T$ is a partition of the leaves into two non-empty
blocks, $A$ and $B$. Removing an edge $e$ from a phylogenetic
$X$-tree $T$ divides $T$ into two connected components, which
induces a split of the leaf set $X$. We will call this the split
associated with $e$. The collection of all the splits associated
with the edges of $T$ is called the splits of $T$ denoted by
$\mathcal S(T)$. Two splits $A_1|B_1$ and $A_2|B_2$ of $X$ are
compatible if at least one of the four intersections $A_1\cap
A_2,A_1\cap B_2, B_1\cap A_2,$ and $B_1\cap B_2$ is empty. Also note
that a collection $\mathcal S$ of splits of $X$ is compatible if it
is contained in the splits of some tree $T$ (\cite{Bu}). We adopt
all of these notations in \cite{Br}.

\begin{thm}[\cite{PS1}]
A collection $\mathcal S$ of splits of $X$ is pairwise compatible if
and only if there exists a tree $T$ such that $\mathcal S=\mathcal
S(T)$.
\end{thm}

Let $X=[n]:=\{1,\cdots,n\}$ and $m$ be the number of states in the
alphabet, $$m=|\Sigma|=\begin{cases} 2,\quad \Sigma=\{0,1\}\\4,\quad
\Sigma=\{A,C,G,T\}.\end{cases}$$ Set $p_{i_1\cdots i_n}$ is the
joint probability that leaf $j$ is observed to be in state $i_j$ for
all $j\in\{1,\cdots,n\}$. Write $P$ for the entire probability
distribution.

\begin{defn} A flattening along a partition $A|B$ is the $m^{|A|}$
by $m^{|B|}$ matrix where the rows are indexed by the possible
states for the leaves in $A$ and the columns are indexed by the
possible states for the leaves in $B$. The entries of this matrix
are given by the joint probabilities of observing the given pattern
at the leaves. We write $\mbox{Flat}_{A|B}(P)$ or shortly
$F_{A|B}(P)$ for this matrix.
\end{defn}

Next we define a measurement that a general partition of the leaves
is close to a split. If $A$ is a subset of the leaves of $T$, then
let $T_A$ be the subtree induced by the leaves in $A$. That is,
$T_A$ is the minimal set of edges needed to connect the leaves in
$A$.
\begin{defn} Suppose that $A|B$ is a partition of $[n]$. The
distance between the partition $A|B$ and the nearest split, written
$e(A,B)$, is the number of edges that occur in $T_A\cap T_B$.
\end{defn}

Notice that $e(A,B) = 0$ exactly when $A|B$ is a split. Consider
$T_A \cap  T_B$ as a subtree of $T_A$. Color the nodes in $T_A \cap
T_B$ red, the nodes in $T_A \setminus (T_A \cap T_B)$ blue. Say that
a node is monochromatic if it and all of its neighbors are of the
same color. We let $\text{mono}(A)$ be the number of monochromatic
red nodes.

\begin{defn} Define $\text{mono}(A)$ as the number of nodes in $T_A\cap  T_B$
that do not have a node in $T_A \setminus (T_A \cap T_B)$ as a
neighbor.
\end{defn}

The following theorem shows how close a partition is to being a
split with the rank of the flattening associated to that partition.
Originally this theorem is proved for the case that $T$ is a
trivalent tree. On the other hand, we have the same result for the
non-trivalent tree $T$ whose proof is almost same as original one in
\cite{Er}.

\begin{thm}
\label{thm2.5} Let $A|B$ be a partition of $[n]$, $T$ be an unrooted
tree which is not necessarily trivalent with leaves labeled by
$[n]$, and assume that the joint probability distribution $P$ comes
from a Markov model on $T$ with an alphabet with $m$ letters. Then
the generic rank of the flattening $\mbox{F}_{A|B}(P)$ is given by
$$\min
(m^{e(A,B)+1-\mbox{mono}(A)},m^{e(A,B)+1-\mbox{mono}(B)},m^{|A|},m^{|B|}).$$
\end{thm}
\begin{proof}
Refer to \cite{Er}, Theorem 19.5.
\end{proof}

Using Theorem \ref{thm2.5} we have the following corollaries.

\begin{cor}\label{cor2.6}
If $A|B$ is a split in the tree, the generic rank of $F_{A|B}(P)$ is
m.
\end{cor}

\begin{cor}\label{cor2.7}
If $A|B$ is not a split in a trivalent tree and we have $|A|, |B|
\ge 2$ then the generic rank of $\mbox{F}_{A|B}(P)$ is at least
$m^2$.
\end{cor}

For the non-trivalent tree case we have a different result comparing
to Corollary \ref{cor2.7}, i.e., generic rank of ${F}_{A|B}(P)$ is
at least $m$. The reason for the different result comes from
considering the following 4-valent tree with $A=\{1,2\},
B=\{3,4,5,6,7,8\}$. Actually, in this case we have $R=\{v_1\}$,
$e(A,B)+1-\mbox{momo}(A)=1, e(A,B)+1-\mbox{momo}(B)=1$ for non-split
$A|B$. Hence we get $\mbox{rank}F_{A|B}(P)= m$.

\begin{figure}
\begin{picture}(200,110)

\put(10,110){$1$}

\put(0,70){$2$}

\put(10,30){$3$}

\put(50,70){\line(-1,0){40}}

\put(50,70){\line(-1,-1){30}}

\put(50,70){\line(-1,1){30}}

\put(50,70){\line(1,0){30}}

\put(50,75){$v_1$}

\put(70,75){$v_2$}

\put(80,70){\line(1,1){60}}

\put(100,105){$v_3$}

\put(110,100){\line(1,-1){30}}

\put(145,60){$5$}

\put(145,130){$4$}

\put(80,70){\line(1,-1){30}}

\put(110,45){$v_4$}

\put(110,40){\line(-1,-1){30}}

\put(70,0){$7$}

\put(110,40){\line(1,0){40}}

\put(140,30){$v_5$}

\put(150,40){\line(1,1){30}}

\put(185,75){$6$}

\put(150,40){\line(1,-1){30}}

\put(185,5){$8$}

\end{picture}

\caption{Non-trivalent tree}

\end{figure}
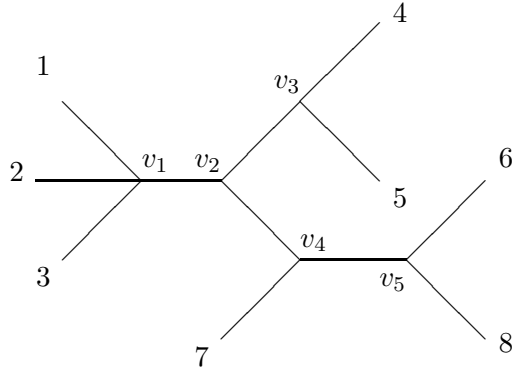

A singular value decomposition of a $m\times n$ matrix $A$ (with $m
\ge n$) is a factorization $A = U\Sigma V^T$ where $U$ is $m\times
n$ and satisfies $U^TU = I$, $V$ is $n \times n$ and satisfies $V^TV
= I$ and $\Sigma = \mbox{diag}(\sigma_1, \sigma_2, \cdots ,
\sigma_n)$, where $\sigma_1 \ge \sigma_2 \ge \cdots \ge \sigma_n \ge
0$ are called the singular values of $A$.


We need the following theorem to define svd distance in the next section.
\begin{thm}[\cite{D}, Theorem 3.3]\label{thm:svd-dis}
The distance from $A$ to the nearest rank $k$ matrix is
$\displaystyle\min
_{\mbox{rank}(B)=k}\|A-B\|_F=\sqrt{\sum^n_{i=k+1}\sigma^2_i}$ in the
Frobenius norm.
\end{thm}

\section{Algorithm for constructing a phylogenetic tree}

In this section, we have an algorithm for constructing a
phylogenetic tree using SVD of flattenings which improves Erikkson's
algorithm in view of numerical stability (cf. \cite{Er}). First we
define a function called svd-cherry as follows.

\begin{defn}\label{SVD-dis}
For each distinct pair $(s_i,s_j)$ in the $k$
species $\mS_k=\{s_1, s_2,\cdots, s_k\}$,
svd distance $d_F(s_i,s_j,\mS_k)$ between $s_i$, and $s_j$ in $\mS_k$ is
defined by the distance from the flattening $F_{\{s_i,s_j\}|
\{s_1,\cdots,s_k\}\backslash\{s_i,s_j\}}(P)$ to the nearest rank $m$
matrix in the Frobenius norm.
\end{defn}

\begin{defn}\label{SVD-cherry}
For given $k$ species  $\mS_k=\{s_1, s_2,\cdots, s_k\}$,
 Define $$\text{\rm svd-cherry}(s_1,
s_2,\cdots, s_k):=(s_{i^*},s_{j^*},v)$$ so that
the pair $(s_{i^*},s_{j^*})$ in $\mS_k$ and their svd distance $v$ in
$\mS_k$
 satisfies that
$$v:=d_F(s_{i^*},s_{j^*},\mS_k)=\displaystyle\min\{d_F(s_i,s_j,\mS_k)~|~(i,j)\\ \in
[k]\times [k],~ i\ne j\}.$$
\end{defn}

Using Definition \ref{SVD-cherry}, we have the following
algorithm.\\

\noindent{\bf Algorithm 1 (Building a phylogenetic tree using SVD of
flattenings)}

\noindent {\bf Input}: A multiple alignment of genomic data from $n$
species from the alphabet $\Sigma$ with $m$ states.

\noindent {\bf Output}: An unrooted phylogenetic tree $T$ with $n$
leaves labeled by the species.

\noindent {\bf Initialization}: Partition $n$ species $s_1,
s_2,\cdots, s_n$ into $n$ singletons as $C_{1,1}, C_{1,2}, \cdots,
C_{1,n}$.

\noindent {\bf Loop}: For $k$ from $1$ to $n-3$, perform the
following steps.

{\bf Step 1}:
For each $n-k+1$ species $s_1, s_2,\cdots, s_{n-k+1}$
where $s_l\in C_{k,l}$ is a
representative of $C_{k,l}$, $1\le l\le n-k+1$, find a distinct pair
of clusters $(C_{k,i^*},  C_{k,j^*})$ such that  $\text{\rm svd-cherry}(s_1,
s_2,\cdots, s_{n-k+1}):=(s_{i^*},s_{j^*},v)$ for $s_{i^*}\in C_{k,i^*},s_{j^*}\in C_{k,j^*}$.

{\bf Step 2}:
Choose the pair of clusters $(C_{k,i^{\diamond}},  C_{k,j^{\diamond}})$
which occurs most frequently in Step 1.

{\bf Step 3}: Join $C_{k,{i^{\diamond}}}$ and $C_{k,{j^{\diamond}}}$ together in the tree and
consider this as a new cluster $C_{k+1,1}$. After that rename the
remaining $C_{k,l}$'s as $C_{k+1,2},\cdots, C_{k+1,n-k}$.\\

\begin{Prop}\label{prop3.1}
Algorithm 1  needs to compute SVD at most
$\sum^n_{k=4}[(\frac{n}{k})^k]{k\choose 2}$ times. Here
$[(\frac{n}{k})^k]$ is the maximum possible integer not greater than
$(\frac{n}{k})^k$.
\end{Prop}
\begin{proof}
If we have $k$ clusters $C_1,C_2,\cdots, C_k$, then the number of
possible representatives for each clusters is $\prod^k_1|C_i|$.
Since $\sum^k_1|C_i|=n,$ $\prod^k_1|C_i|$ has maximum
$(\frac{n}{k})^k$ where $|C_1|=|C_2|=\cdots=|C_k|=\frac{n}{k}$. Here
$|C_i|$ is the cardinality of $C_i$. Thus we manipulate SVD at most
$\sum^n_{k=4}[(\frac{n}{k})^k]{k\choose 2}$ times in total for the
flattenings with fixed number of rows, i.e., $m^2$.
\end{proof}

Each cluster $C_{k,l},~ 1\le k\le n-2 ,~ 1\le l < n-k+1$ in Algorithm 1
means a split in the tree.
In Algorithm 1 we have the following hierarchy of $C_{k,l}$'s.
In the Initialization, $C_{1,l} ~(1\le l\le n)$ mean $n$ trivial
splits, in other words, outer edges in tree
$T$. At the end of the first loop, there is one new cluster
among  $C_{2,l}~ (1\le l\le n-1$), which means  one new
split in $T$. At the end of each $k$-th loop from $k=1$ up to $k=n-3$, we
obtain one new edge in $T$. In total we have the exact $2n-3$
splits in $T$.
\begin{eqnarray*}
&&k=1:~~\quad\qquad \qquad~~~C_{1,1}~~C_{1,2}~~C_{1,3}~~\cdots\cdots\cdots~~C_{1,n-2}~~C_{1,n-1}~~C_{1,n}\\
&&k=2:~~\quad\qquad \qquad~~~C_{2,1}~~C_{2,2}~~C_{2,3}~~\cdots\cdots\cdots~~C_{2,n-2}~~C_{2,n-1}\\
&&k=3:~~\quad\qquad \qquad~~~C_{3,1}~~C_{3,2}~~C_{3,3}~~\cdots\cdots\cdots~~C_{3,n-2}\\
&&\qquad\quad\quad\qquad\qquad\vdots\qquad\vdots\qquad\vdots\qquad\vdots~~\\
&&k=n-3:\quad\quad C_{n-3,1}~~C_{n-3,2}~~C_{n-3,3}~~C_{n-3,4}\\
&&k=n-2:\quad\quad C_{n-2,1}~~C_{n-2,2}~~C_{n-2,3}
\end{eqnarray*}

The matrix size $m^2\times
m^{k-2}$ of flattenings may be large where $k$ varies from $n$ to 4, on the
other hand flattenings are very sparse.
Thus, it is faster to  compute the eigenvalues of $A^TA$ of fixed
size $m^2\times m^2$ for every flattening $A$ than singular values
of $A$ itself of size $m^2\times m^{k-2}$. Erikkson computed
singular values of flattenings of various huge size $m^{|A|}\times
m^{|B|}$ where $A|B$ is a partition of $[n]$. That must cause
numerical instability. We, however, avoid computational difficulties
which come from numerical instability since we only deal with $A^TA$
of fixed size $m^2\times m^2$ for every flattening $A$. Although we
also have difficulty in computing lots of SVD of matrices of fixed
size $m^2\times m^2$, if the number of species grows, Algorithm 1 is
fit for parallel computing, especially, grid computing which
arranges lots of volunteer computing resources to do distributed
computing.
\begin{thm}\label{thm3.2}
Algorithm 1 is statistically consistent.
\end{thm}
\begin{proof}
By Corollary \ref{cor2.6}, we can see that $d_F(s_i,s_j,\mS_k)$ goes
to 0 if $\{s_i,s_j\}| \{s_1,\cdots,s_k\}\\\backslash \{s_i,s_j\}$ is
a true split of $\mS_k=\{s_1,\cdots,s_k\}$. While Corollary
\ref{cor2.7} shows that $d_F(s_i,s_j,\mS_k)$ does not go to 0 if
$\{s_i,s_j\}| \{s_1,\cdots,s_k\}\backslash \{s_i,s_j\}$ is a
partition which is not a split of $\mS_k$. Hence, as the empirical
distribution approaches the true one, the distance of a split from
rank $m$ will go to zero while the distance from rank $m$ of a
non-split will not. Therefore Algorithm 1 picks a correct split at
each loop.
\end{proof}

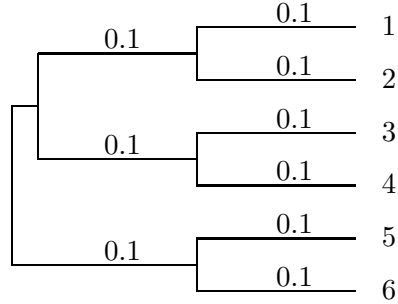
\begin{figure}
\begin{picture}(300,110)

\put(240,102){$1$}

\put(240,82){$2$}

\put(240,62){$3$}

\put(240,42){$4$}

\put(240,22){$5$}

\put(240,2){$6$}

\put(200,107){$0.1$}

\put(200,87){$0.1$}

\put(200,67){$0.1$}

\put(200,47){$0.1$}

\put(200,27){$0.1$}

\put(200,7){$0.1$}

\put(135,97){$0.1$}

\put(135,57){$0.1$}

\put(135,17){$0.1$}

\put(170,105){\line(1,0){60}}

\put(170,105){\line(0,-1){20}}

\put(170,85){\line(1,0){60}}

\put(170,65){\line(1,0){60}}

\put(170,65){\line(0,-1){20}}

\put(170,45){\line(1,0){60}}

\put(170,25){\line(1,0){60}}

\put(170,25){\line(0,-1){20}}

\put(170,5){\line(1,0){60}}

\put(110,95){\line(1,0){60}}

\put(110,95){\line(0,-1){40}}

\put(110,55){\line(1,0){60}}

\put(110,75){\line(-1,0){10}}

\put(100,75){\line(0,-1){60}}

\put(100,15){\line(1,0){70}}

\end{picture}

\caption{Model tree for 6 species.} \label{ex1}

\end{figure}

\noindent {\bf Example}

We begin with an alignment of DNA data of length 1000 for 6 species,
labeled $1, \cdots, 6,$ simulated from the tree in Figure \ref{ex1}
with all branch lengths equal to 0.1.
For the loop ($k=1$), let
$C_{1,1}=\{1\},C_{1,2}=\{2\},C_{1,3}=\{3\},C_{1,4}=\{4\},C_{1,5}=\{5\},C_{1,6}=\{6\}$
and consider all pairs of the 6 species. In the following results, the
svd-val($s_1,s_2|s_3,\cdots,s_k$) is the distance
 from the flattening $F_{\{s_1,s_2\}|
\{s_3,\cdots,s_k\}}(P)$ to the nearest rank $4$ matrix in the
Frobenius norm as in Theorem \ref{thm:svd-dis}
\texttt{
\begin{center}
 loop: k = 1\\
 svd-val(  1,  2 |  3,  4,  5,  6 ) = 0.0188\\
 svd-val(  1,  3 |  2,  4,  5,  6 ) = 0.2102\\
 svd-val(  1,  4 |  2,  3,  5,  6 ) = 0.2103\\
 svd-val(  1,  5 |  2,  3,  4,  6 ) = 0.4297\\
 svd-val(  1,  6 |  2,  3,  4,  5 ) = 0.4298\\
 svd-val(  2,  3 |  1,  4,  5,  6 ) = 0.2095\\
 svd-val(  2,  4 |  1,  3,  5,  6 ) = 0.2096\\
 svd-val(  2,  5 |  1,  3,  4,  6 ) = 0.4298\\
 svd-val(  2,  6 |  1,  3,  4,  5 ) = 0.4297\\
 svd-val(  3,  4 |  1,  2,  5,  6 ) = 0.0128\\
 svd-val(  3,  5 |  1,  2,  4,  6 ) = 0.4779\\
 svd-val(  3,  6 |  1,  2,  4,  5 ) = 0.4779\\
 svd-val(  4,  5 |  1,  2,  3,  6 ) = 0.4779\\
 svd-val(  4,  6 |  1,  2,  3,  5 ) = 0.4779\\
 svd-val(  5,  6 |  1,  2,  3,  4 ) = 0.0076\\
 min value cherry = ( 5, 6 )
\end{center}
}
After first loop, since
svd-cherry($1,2,3,4,5,6$)=(5,~6,~0.0076), we have a new cluster $C_{2,1}=\{5,6\}$
and rename $C_{2,2}=\{1\},C_{2,3}=\{2\},C_{2,4}=\{3\}, C_{2,5}=\{4\}$.\\
\texttt{ \begin{center}
 loop: k = 2\\
 svd-val(  1,  2 |  3,  4,  5 ) = 0.0216\\
 svd-val(  1,  3 |  2,  4,  5 ) = 0.2104\\
 svd-val(  1,  4 |  2,  3,  5 ) = 0.2105\\
 svd-val(  1,  5 |  2,  3,  4 ) = 0.0214\\
 svd-val(  2,  3 |  1,  4,  5 ) = 0.2094\\
 svd-val(  2,  4 |  1,  3,  5 ) = 0.2095\\
 svd-val(  2,  5 |  1,  3,  4 ) = 0.0290\\
 svd-val(  3,  4 |  1,  2,  5 ) = 0.0127\\
 svd-val(  3,  5 |  1,  2,  4 ) = 0.2101\\
 svd-val(  4,  5 |  1,  2,  3 ) = 0.2099\\
 min value cherry = ( 3, 4 )\\
 --------------------------------------------------\\
 svd-val(  1,  2 |  3,  4,  6 ) = 0.0169\\
 svd-val(  1,  3 |  2,  4,  6 ) = 0.2090\\
 svd-val(  1,  4 |  2,  3,  6 ) = 0.2091\\
 svd-val(  1,  6 |  2,  3,  4 ) = 0.0213\\
 svd-val(  2,  3 |  1,  4,  6 ) = 0.2084\\
 svd-val(  2,  4 |  1,  3,  6 ) = 0.2085\\
 svd-val(  2,  6 |  1,  3,  4 ) = 0.0257\\
 svd-val(  3,  4 |  1,  2,  6 ) = 0.0127\\
 svd-val(  3,  6 |  1,  2,  4 ) = 0.2090\\
 svd-val(  4,  6 |  1,  2,  3 ) = 0.2089\\
 min value cherry=( 3, 4 )\\
\end{center}
}
For the loop ($k=2$), first take 5 as a representative
of $C_{2,1}=\{5,6\}$ and get svd-cherry($1,2,3,4,5$)
=(3,~4,~0.0127). Next choose 6 as a representative of
$C_{2,1}=\{5,6\}$ and get svd-cherry($1,2,3,4,6$)
=(3,~4,~0.0127). Most frequent pair of clusters in
$C_{2,1},\cdots,C_{2,5}$  is $(C_{2,4},C_{2,5})$.  We obtain a new cluster
$C_{3,1}=\{3,4\}$ and rename $C_{3,2}=\{5,6\},~ C_{3,3}=\{1\}, ~C_{3,4}=\{2\}$.\\
\texttt{
\begin{center}
 loop: k = 3\\
 svd-val(  1,  2 |  5,  3 ) = 0.0213\\
 svd-val(  1,  5 |  2,  3 ) = 0.0208\\
 svd-val(  1,  3 |  2,  5 ) = 0.0285\\
 min value cherry=( 1, 5 )\\
 --------------------------------------------------\\
 svd-val(  1,  2 |  6,  3 ) = 0.0167\\
 svd-val(  1,  6 |  2,  3 ) = 0.0207\\
 svd-val(  1,  3 |  2,  6 ) = 0.0252\\
 min value cherry=( 1, 2 )\\
 --------------------------------------------------\\
 svd-val(  1,  2 |  5,  4 ) = 0.0214\\
 svd-val(  1,  5 |  2,  4 ) = 0.0221\\
 svd-val(  1,  4 |  2,  5 ) = 0.0295\\
 min value cherry=( 1, 2 )\\
 --------------------------------------------------\\
 svd-val(  1,  2 |  6,  4 ) = 0.0168\\
 svd-val(  1,  6 |  2,  4 ) = 0.0219\\
 svd-val(  1,  4 |  2,  6 ) = 0.0262\\
 min value cherry=( 1, 2 )\\
\end{center}
}
For the loop ($k=3$), first take 3 as a representative
of $C_{3,1}=\{3,4\}$, 5 as a representative of $C_{3,2}=\{5,6\}$,
then we get svd-cherry($3,5,1,2$)=(1,~5,~0.0208). Next choose 3 as a
representative of $C_{3,1}=\{3,4\}$, 6 as a representative of
$C_{3,2}=\{5,6\}$, get svd-cherry($3,6,1,2$)=(1,~2,~0.0167). By
the same manner we have svd-cherry($5,4,1,2$)=(1,~2,~0.0214),
svd-cherry($6,4,1,2$)=(1,~2,~0.0168).
Most frequent pair of clusters in
$C_{3,1},\cdots,C_{3,4}$  is $(C_{3,3},C_{3,4})$.  We obtain a new cluster
$C_{4,1}=\{1,2\}$ and rename $C_{4,2}=\{5,6\},~ C_{4,3}=\{3,4\}$.
We can join  these
three clusters $C_{4,1}, C_{4,2}, C_{4,3}$ to make an unrooted tree.

\section{Simplified tree constructing algorithm}

In Algorithm 1, if we can reduce the number of feasible
representative of each cluster $C_{k,l}$ using some available a
priori information, then computational cost can be saved. In this
section, for example, we choose the unique feasible representative
of each cluster which has the smallest distance from species
outside the cluster. \\

\noindent{\bf Algorithm 2 (Simplified tree constructing algorithm)}

\noindent {\bf Input}: A multiple alignment of genomic data from $n$
species from the alphabet $\Sigma$ with $m$ states.

\noindent {\bf Output}: An unrooted phylogenetic tree $T$ with $n$
leaves labeled by the species.

\noindent {\bf Initialization}: Partition $n$ species $s_1,
s_2,\cdots, s_n$ into $n$ singletons as $C_{1,1}, C_{1,2}, \cdots,
C_{1,n}$.

\noindent {\bf Loop}: For $k$ from $1$ to $n-3$, perform the
following steps.

{\bf Step 1}: For each cluster $C_{k,j},~ 1\le j \le n-k+1$, choose
the representative $s_j^R\in C_{k,j}$ by  the following;

\noindent \hspace{0.5cm} For all $s\in C_{k,j}$, calculate
$v_s={\sum_{s'\in [n]\backslash C_{k,j}}p(s,s')}$ where $p(s,s')$ is the

\noindent \hspace{0.5cm} proportion of different nucleotides between
two species $s,s'$. Choose $s_j^R\in C_{k,j}$

\noindent \hspace{0.5cm} which has the smallest value $v_s$ for all
$s\in C_{k,j}$.

{\bf Step 2}: For $(s^R_1, s^R_2,\cdots, s^R_{n-k+1})$ where
$s^R_l\in C_{k,l}, 1\le l\le n-k+1$, find a distinct pair of
clusters $(C_{k,i^*},  C_{k,j^*})$ such that  $\text{\rm
svd-cherry}(s_1^R, s_2^R,\cdots, s^R_{n-k+1}):=(s_{i^*},s_{j^*},v)$
for $s_{i^*}\in C_{k,i^*},s_{j^*}\in C_{k,j^*}$.

{\bf Step 3}: Join $C_{k,{i^{*}}}$ and $C_{k,{j^{*}}}$ together in the tree and
consider this as a new cluster $C_{k+1,1}$. After that rename the
remaining $C_{k,l}$'s as $C_{k+1,2},\cdots, C_{k+1,n-k}$.\\

Note that Algorithm 2 is much faster to construct phylogenetic
trees with many leaves since it uses only one representative
for each cluster $C_{k,l}$.

\section{Performance analysis of tree constructing algorithms}

\subsection{Building phylogenetic trees with simulated data}

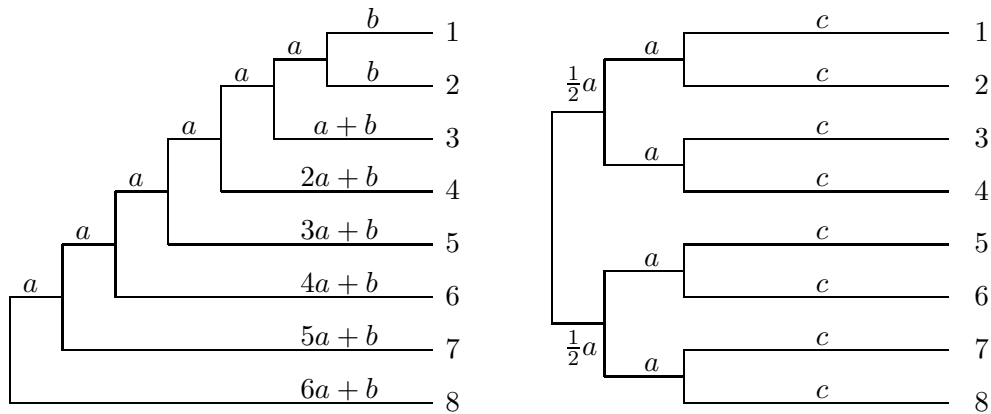
\begin{figure}

\begin{picture}(400,150)

\put(190,142){$1$}

\put(190,122){$2$}

\put(190,102){$3$}

\put(190,82){$4$}

\put(190,62){$5$}

\put(190,42){$6$}

\put(190,22){$7$}

\put(190,2){$8$}

\put(160,147){$b$}

\put(160,127){$b$}

\put(140,107){$a+b$}

\put(135,87){$2a+b$}

\put(135,67){$3a+b$}

\put(135,47){$4a+b$}

\put(135,27){$5a+b$}

\put(135,7){$6a+b$}

\put(145,145){\line(1,0){40}}

\put(145,145){\line(0,-1){20}}

\put(145,125){\line(1,0){40}}

\put(130,137){$a$}

\put(125,135){\line(1,0){20}}

\put(125,135){\line(0,-1){30}}

\put(125,105){\line(1,0){60}}

\put(110,127){$a$}

\put(105,125){\line(1,0){20}}

\put(105,125){\line(0,-1){40}}

\put(105,85){\line(1,0){80}}

\put(90,107){$a$}

\put(85,105){\line(1,0){20}}

\put(85,105){\line(0,-1){40}}

\put(85,65){\line(1,0){100}}

\put(70,87){$a$}

\put(65,85){\line(1,0){20}}

\put(65,85){\line(0,-1){40}}

\put(65,45){\line(1,0){120}}

\put(50,67){$a$}

\put(45,65){\line(1,0){20}}

\put(45,65){\line(0,-1){40}}

\put(45,25){\line(1,0){140}}

\put(30,47){$a$}

\put(25,45){\line(1,0){20}}

\put(25,45){\line(0,-1){40}}

\put(25,5){\line(1,0){160}}


\put(390,142){$1$}

\put(330,147){$c$}

\put(265,137){$a$}

\put(265,97){$a$}

\put(265,57){$a$}

\put(265,17){$a$}

\put(235,123){$\frac{1}{2}a$}

\put(235,23){$\frac{1}{2}a$}

\put(280,145){\line(1,0){100}}

\put(280,145){\line(0,-1){20}}

\put(280,125){\line(1,0){100}}

\put(250,135){\line(1,0){30}}

\put(250,135){\line(0,-1){40}}

\put(250,95){\line(1,0){30}}

\put(230,115){\line(1,0){20}}

\put(230,115){\line(0,-1){80}}


\put(230,35){\line(1,0){20}}

\put(230,75){\line(0,-1){20}}

\put(390,122){$2$}

\put(330,127){$c$}

\put(390,102){$3$}

\put(330,107){$c$}

\put(280,105){\line(1,0){100}}

\put(280,105){\line(0,-1){20}}

\put(280,85){\line(1,0){100}}

\put(390,82){$4$}

\put(330,87){$c$}

\put(390,62){$5$}

\put(330,67){$c$}

\put(280,65){\line(1,0){100}}

\put(280,65){\line(0,-1){20}}

\put(280,45){\line(1,0){100}}

\put(250,55){\line(1,0){30}}

\put(250,55){\line(0,-1){40}}

\put(250,15){\line(1,0){30}}

\put(390,42){$6$}

\put(330,47){$c$}

\put(390,22){$7$}

\put(330,27){$c$}

\put(280,25){\line(1,0){100}}

\put(280,25){\line(0,-1){20}}

\put(280,5){\line(1,0){100}}

\put(390,2){$8$}

\put(330,7){$c$}

\end{picture}

\caption{Model trees (A)(left) and (B)(right).} \label{Model trees}

\end{figure}

We chose  phylogenetic tree models as in Figure \ref{Model trees}
and simulated DNA sequence data on these trees using the program
seq-gen (\cite{RG}). Figure \ref{Model trees} shows variables $a, b,
c$ in the trees. These trees were chosen as difficult trees in
\cite{SN}. Next, we built trees using Algorithm 1,2 and neighbor
joining algorithm with Jukes-Cantor distance from these data,
respectively. For each algorithm, we plotted percent of tree
reconstructed among 1000 DNA data set for various sequence lengths.

\begin{figure}
\subfigure{{\epsfig{file=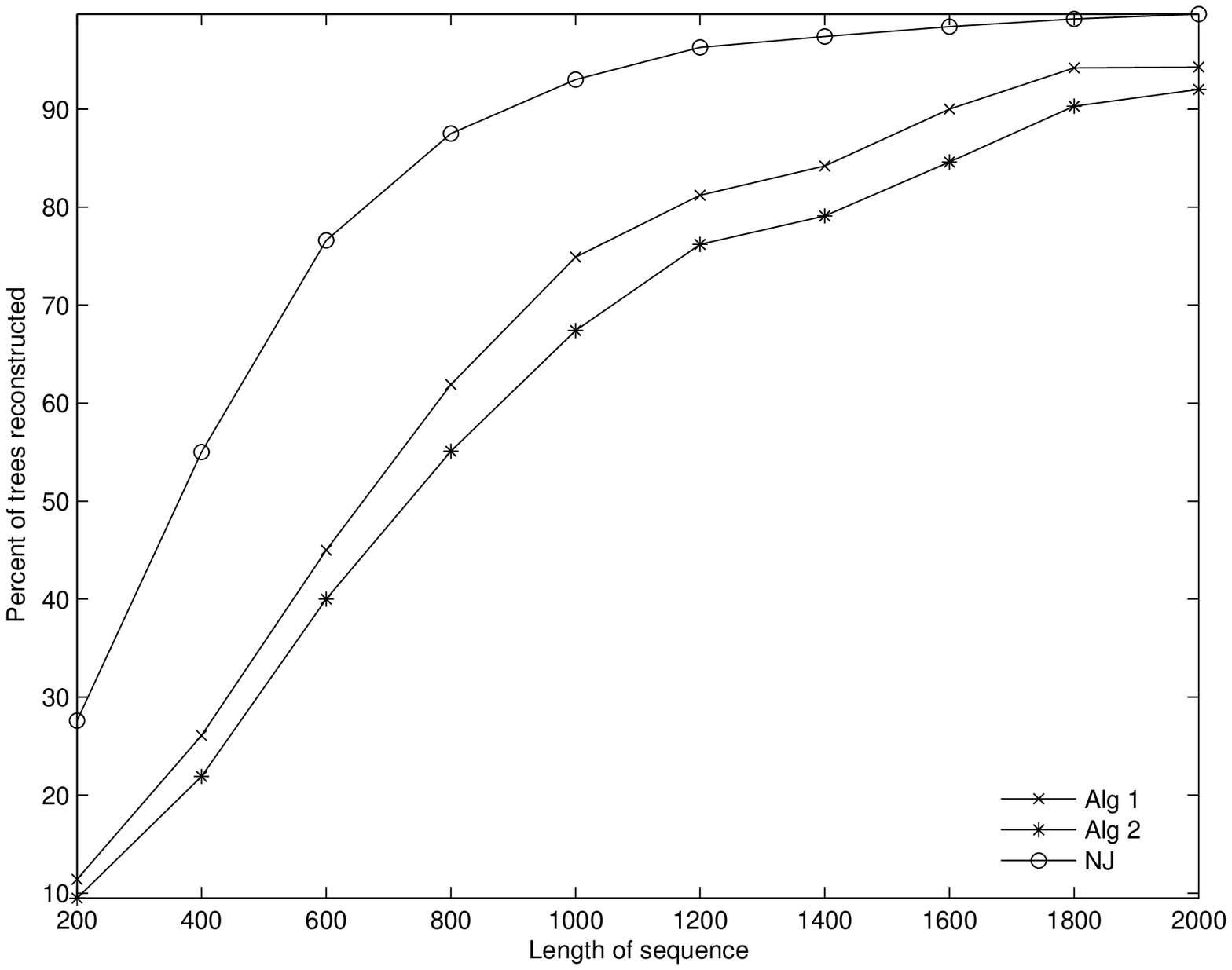,height=5cm,width=6.5cm}}}
\subfigure{{\epsfig{file=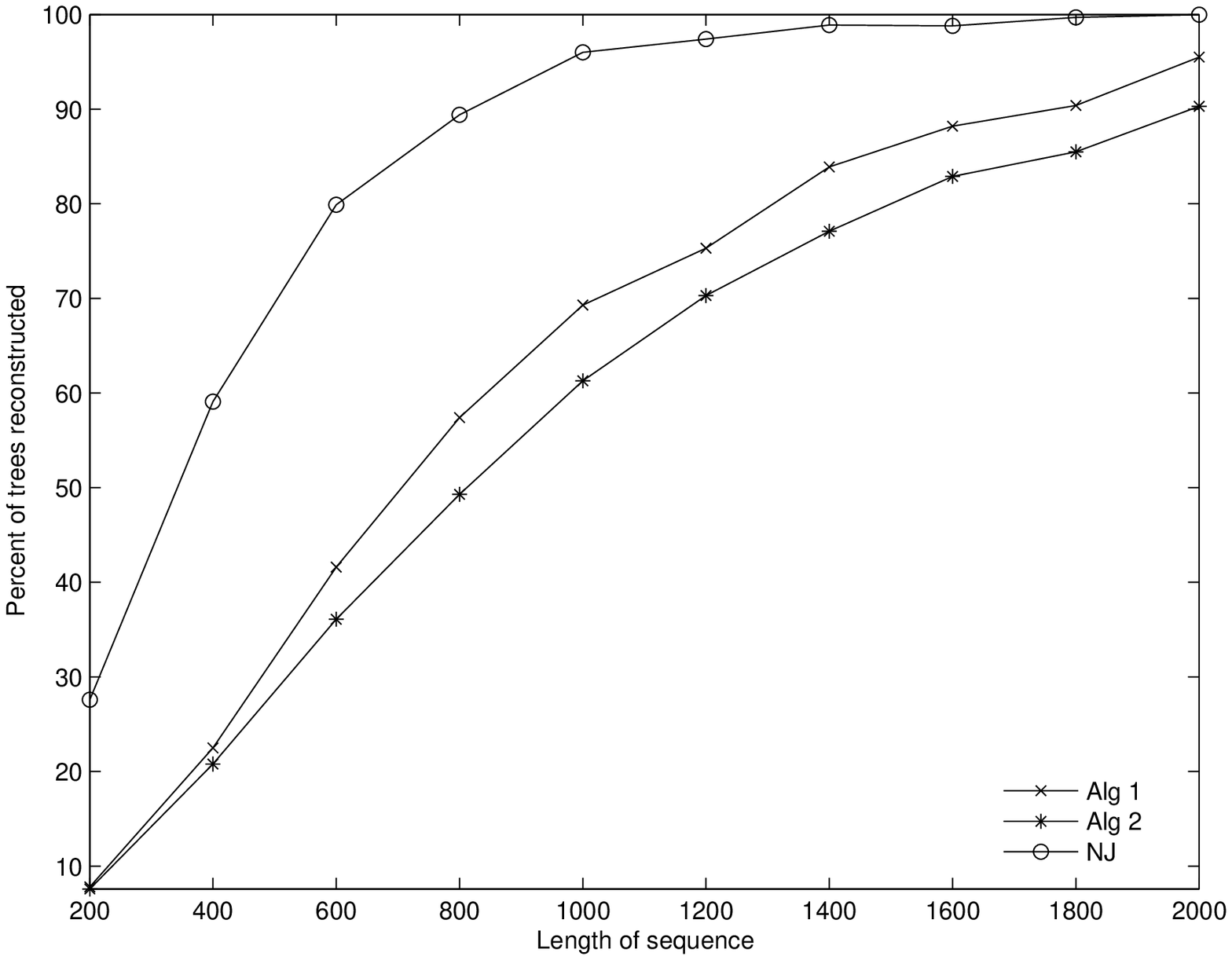,height=5cm,width=6.5cm}}}
\caption{Simulation results of tree construction methods for Model
trees A(left) and B(right) in the case $a=.01,~b=.04, ~c=.07$.}
\label{c=.07}
\end{figure}
\begin{figure}
\subfigure{{\epsfig{file=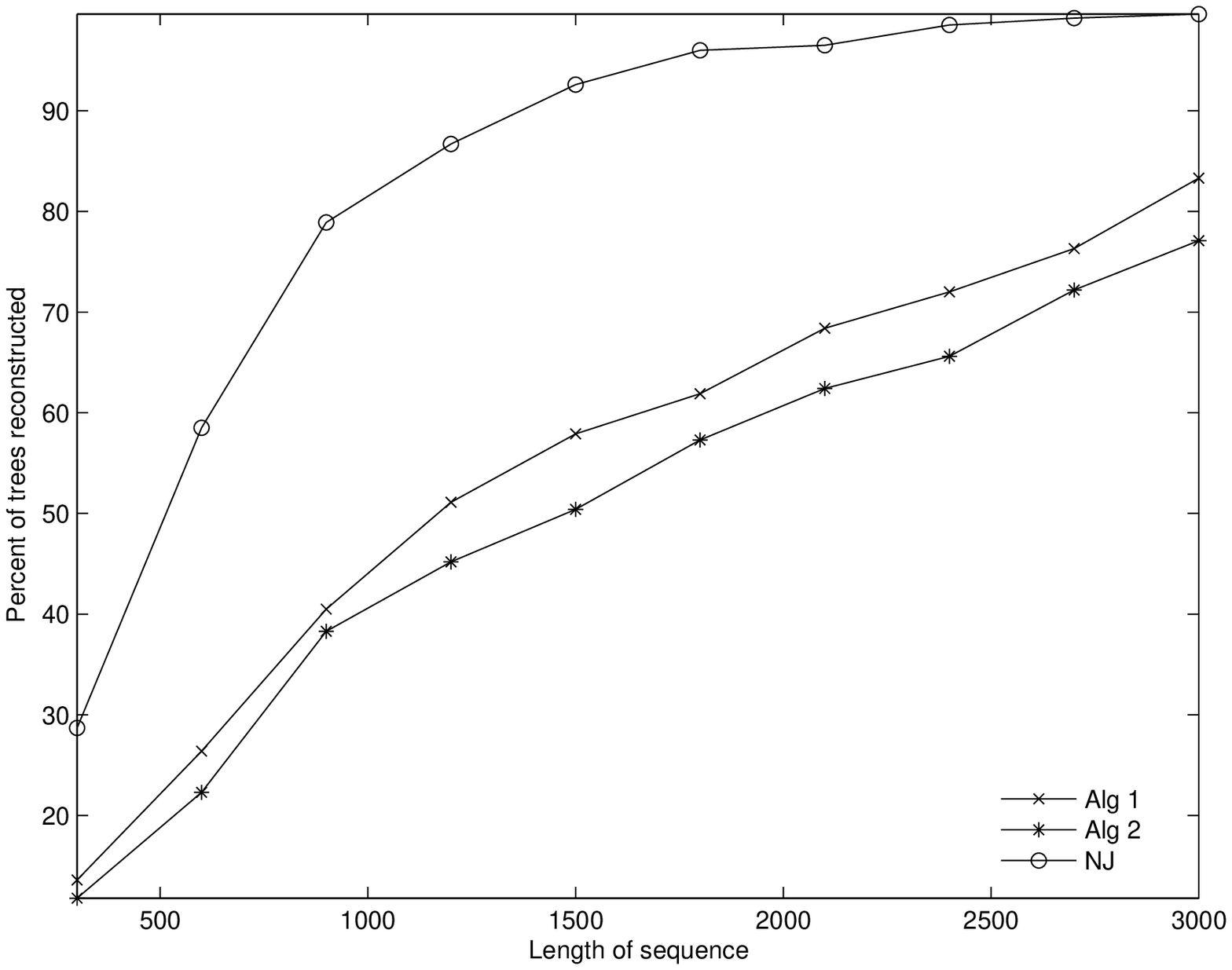,height=5cm,width=6.5cm}}}
\subfigure{{\epsfig{file=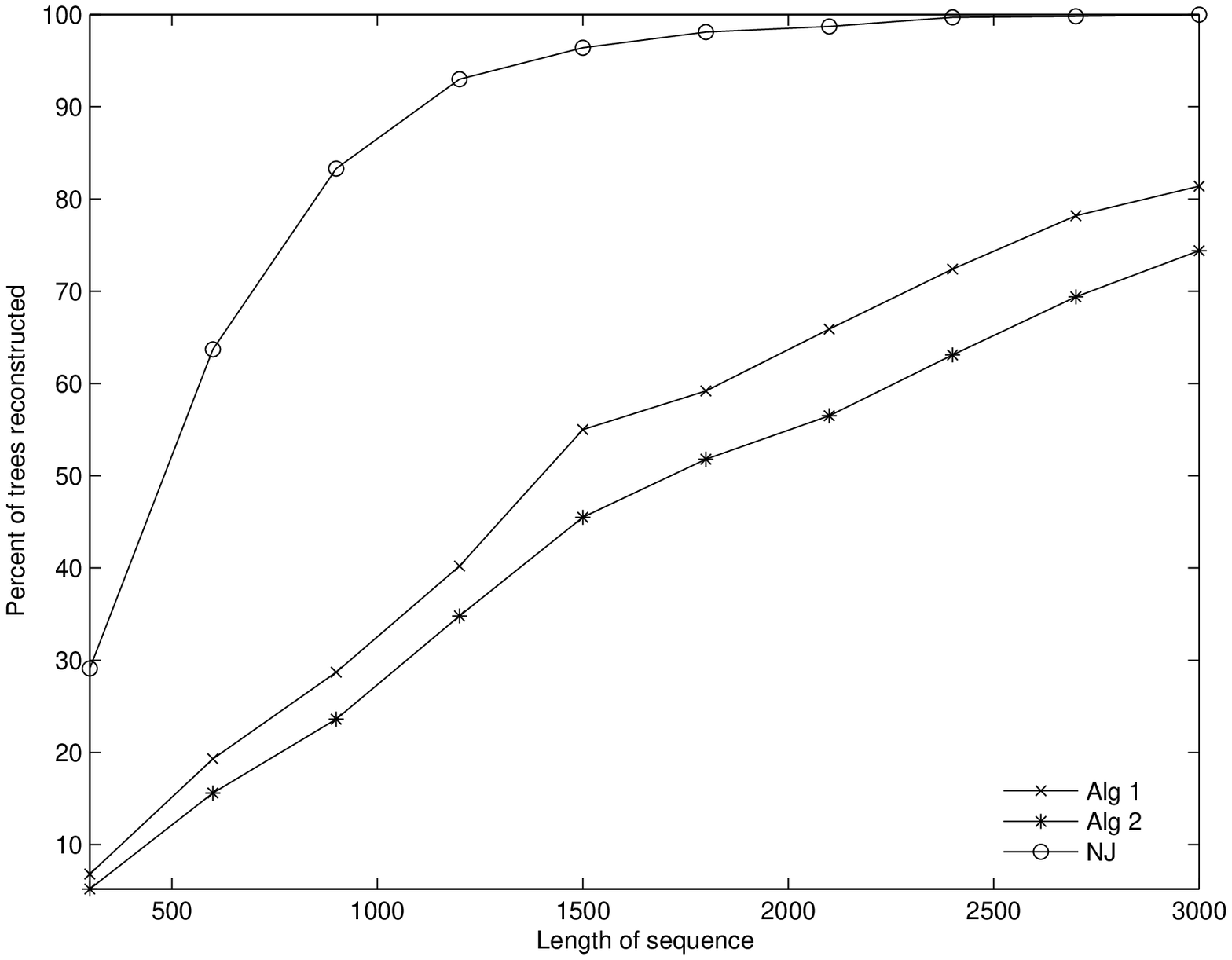,height=5cm,width=6.5cm}}}
\caption{Simulation results of tree construction methods for Model
trees A(left) and B(right) in the case $a=.02,~b=.13, ~c=.19$.}
\label{c=.19}
\end{figure}

Figure \ref{c=.07} shows the results for the case of $a=.01, b=.04,
c=.07$ for both model trees in Figure \ref{Model trees}. The results
for the case of $a=.02, b=.13, c=.19$ are shown in Figure
\ref{c=.19}. Algorithm 1 shows better performance than Algorithm 2,
but, worse than the neighbor joining algorithm. It might be expected
because the used DNA data were simulated by distance based algorithm.

\begin{figure}
{\epsfig{file=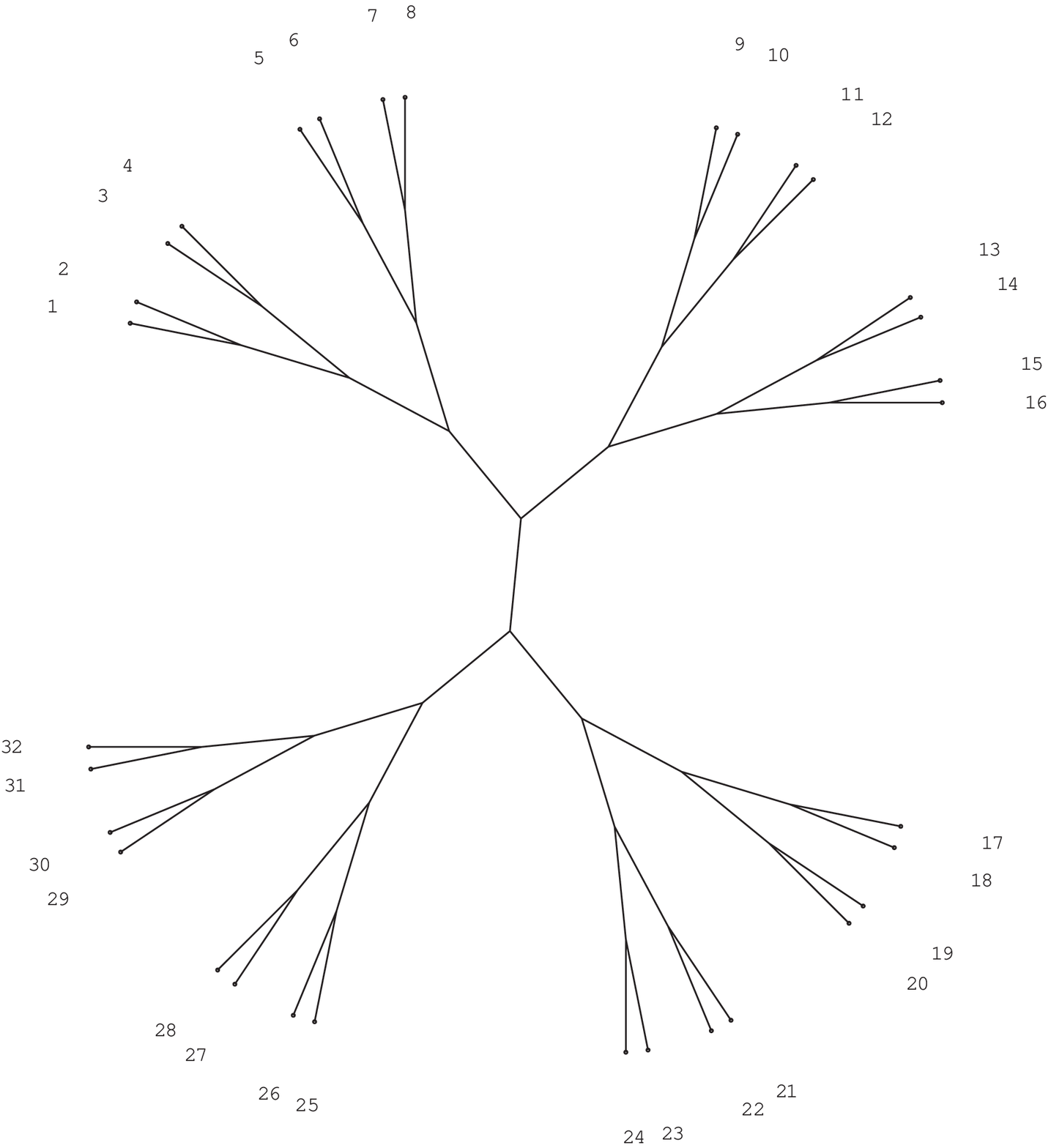,height=10cm,width=10cm}}
\caption{Phylogenetic tree with 32 leaves whose all of the edges have
branch length .1.} \label{32taxa}
\end{figure}

We tested Algorithm 1,2 to reconstruct a tree with many species, for
example 32 species. We simulated 100 DNA data sets of length 1000 for
32 species from the tree in Figure \ref{32taxa} with all branch
lengths equal to .1. We got 94 \% of reconstruction rate using
Algorithm 2, whereas 99 \% of reconstruction rate with neighbor
joining algorithm. For the Algorithm 1, we tested only 1 data set
using parallel cluster machine with 10 Athlon 2600 CPUs. It took
about 3 hours for 1 data set. The loop number which took longest
time was 19. Algorithm 1 did $2^{16} {14\choose 2}$ SVD computations
in 19-th loop. The important point is that we change the type of
difficulty in dealing with rebuilding tree of many species from time
and numerical instability  to time only. Furthermore, the difficulty
in time can be overcome in various ways.

\subsection{Building
phylogenetic trees with real data}

For data, we use the September 2005 freeze of the ENCODE alignments.
We restrict our attention to the problem of constructing
phylogenetic tree for 8 species: human, chimp, galago, mouse, rat,
cow, dog, and chicken, which is called rodent problem. We processed
each of the 44 ENCODE regions to obtain data sets which have
ungapped columns greater than 100 bps in length. We obtain 75 data
sets in manually chosen 14 Enm regions and 301 data sets in all 44
Encode regions.

Recall that the Robinson-Foulds metric which is also called the
partition metric was proposed by \cite{RF} is one of the
simplest metrics on trees. The distance between two trees $T_1$
and $T_2$ is defined by
$$d_{RF}(T_1,T_2)=\frac{1}{2}(|\mathcal S(T_1)-\mathcal S(T_2)|+
|\mathcal S(T_2)-\mathcal S(T_1)|).$$ Here $\mathcal S(T)$ is the set of
splits in $T$ and $|\mathcal
S(T_i)-\mathcal S(T_j)|$ is the cardinality of the set $\mathcal
S(T_i)-\mathcal S(T_j)$.
The symmetric distance $d_s(T_1,T_2)$ is twice of
$d_{RF}(T_1,T_2)$.

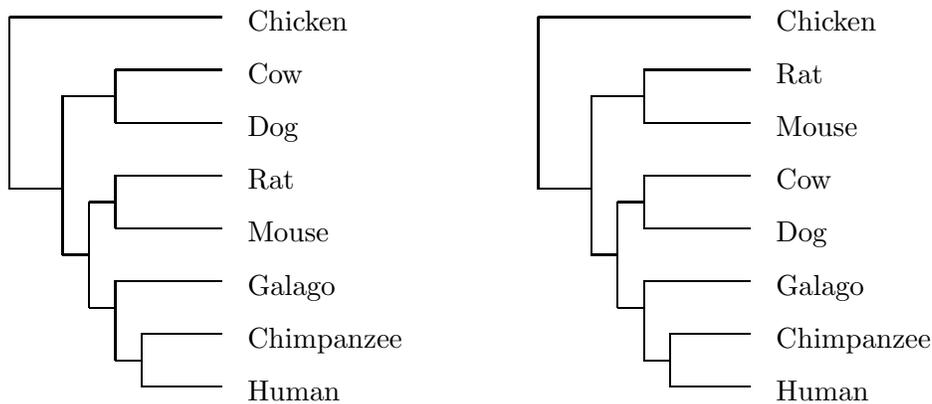
\begin{figure}

\begin{picture}(400,150)

\put(25,145){\line(1,0){80}}

\put(25,145){\line(0,-1){65}}

\put(25,80){\line(1,0){20}}

\put(65,125){\line(1,0){40}}

\put(65,125){\line(0,-1){20}}

\put(45,115){\line(1,0){20}}

\put(45,115){\line(0,-1){60}}

\put(45,55){\line(1,0){10}}

\put(65,105){\line(1,0){40}}

\put(65,85){\line(1,0){40}}

\put(65,85){\line(0,-1){20}}

\put(65,65){\line(1,0){40}}

\put(55,75){\line(1,0){10}}

\put(55,75){\line(0,-1){40}}

\put(55,35){\line(1,0){10}}

\put(65,45){\line(1,0){40}}

\put(65,45){\line(0,-1){30}}

\put(65,15){\line(1,0){10}}

\put(75,25){\line(1,0){30}}

\put(75,25){\line(0,-1){20}}

\put(75,5){\line(1,0){30}}

\put(115,140){Chicken}

\put(115,120){Cow}

\put(115,100){Dog}

\put(115,80){Rat}

\put(115,60){Mouse}

\put(115,40){Galago}

\put(115,20){Chimpanzee}

\put(115,0){Human}

\put(225,145){\line(1,0){80}}

\put(225,145){\line(0,-1){65}}

\put(225,80){\line(1,0){20}}

\put(265,125){\line(1,0){40}}

\put(265,125){\line(0,-1){20}}

\put(245,115){\line(1,0){20}}

\put(245,115){\line(0,-1){60}}

\put(245,55){\line(1,0){10}}

\put(265,105){\line(1,0){40}}

\put(265,85){\line(1,0){40}}

\put(265,85){\line(0,-1){20}}

\put(265,65){\line(1,0){40}}

\put(255,75){\line(1,0){10}}

\put(255,75){\line(0,-1){40}}

\put(255,35){\line(1,0){10}}

\put(265,45){\line(1,0){40}}

\put(265,45){\line(0,-1){30}}

\put(265,15){\line(1,0){10}}

\put(275,25){\line(1,0){30}}

\put(275,25){\line(0,-1){20}}

\put(275,5){\line(1,0){30}}

\put(315,140){Chicken}

\put(315,120){Rat}

\put(315,100){Mouse}

\put(315,80){Cow}

\put(315,60){Dog}

\put(315,40){Galago}

\put(315,20){Chimpanzee}

\put(315,0){Human}

\end{picture}

\caption{Biologically correct tree(left) and the tree which is
obtained by usual algorithm(right).}
\label{rodent problem}

\end{figure}

Rodents have very different morphological features, although their
molecular data is similar to that of the primates. Thus, lots of
biologists pay attention to rodents. Tree construction algorithms
using genomic data usually misplace the rodents, mouse and rat, on
the tree, with respect to other mammals. According to fossil records
and molecular data, we have the biologically correct tree that is
not sure whether it is correct. In this tree we have the primate
clade with human and chimpanzee and then the galago as an outgroup
to these two. The rodent clade (mouse and rat) is a sister group to
the clade (human, chimpanzee and galago) and the clade (dog and cow)
is the outgroup to former 5 species. The chicken is an outgroup to
all of these as a root of this phylogenetic tree. On the other hand,
usual tree reconstruction algorithm mislocate the rodents and so
locate them as an out group to the clade (human, chimpanzee and
galago) (See Figure \ref{rodent problem}).

\begin{table}
\begin{tabular}{|c|c|c|c|c|}
\hline
  method & Algorithm 1 & Algorithm 2 & NJ \\
\hline
  All ($P_c$)& 11.9 & 10.6 & 11.6 \\
\hline
  All ($d_s$)& 2.57 & 2.85 & 2.77\\
\hline
  Enm($P_c$)& 12.0 & 12.0 & 14.6\\
\hline
  Enm ($d_s$)& 2.14 & 2.52 & 2.45\\
\hline
\end{tabular}\vspace{0.5cm}
\caption{Comparing results for algorithms on data from Encode
project. $P_c$ is percent of trees reconstructed and $d_s$ is
symmetric distance.} \label{comparison}
\end{table}

The reasons for this are not entirely known, but it could be since
tree construction methods generally assume the existence of a global
rate matrix for all the species. However, rat and mouse have mutated
faster than the other species. Our algorithms does not assume
anything about the rate matrix (\cite{PS1}, Chapter 21).

In fact, Table \ref{comparison} shows that our algorithms performs
quite well on the ENCODE data sets comparing to NJ(neighbor joining
algorithm with Jukes-Cantor distance) algorithm. Algorithm 1
constructs the correct tree similar to NJ (cf. \cite{Er}, p.357),
but, has shorter symmetric distance $d_s$ on average than NJ
algorithm.

\end{document}